\documentclass[aps,prd,groupedaddress,nofootinbib,superscriptaddress,eqsecnum,amsmath,amssymb]{revtex4}
\usepackage{amsfonts}
\usepackage{graphicx}
\usepackage{bm}
\usepackage{float}
\usepackage[colorlinks=true,
    linkcolor=blue,
    citecolor=blue,
    urlcolor=blue]{hyperref}
\usepackage[figurename=FIG.]{caption}

    \newcommand{\ernst}{\mathcal{E}}
\renewcommand\Re{\operatorname{Re}}
\renewcommand\Im{\operatorname{Im}}

\begin{document}

\title{Ehlers transformations as a tool for constructing accelerating NUT black holes}

\author{Jos\'e Barrientos}
\email{barrientos@math.cas.cz}
\affiliation{Institute of Mathematics of the Czech Academy of Sciences, \v{Z}itn\'a 25, 11567 Praha 1, Czech Republic}
\affiliation{Sede Esmeralda, Universidad de Tarapac\'a, Avenida Luis Emilio Recabarren 2477, Iquique, Chile}

\author{Adolfo Cisterna}
\email{adolfo.cisterna.r@mail.pucv.cl}
\affiliation{Sede Esmeralda, Universidad de Tarapac\'a, Avenida Luis Emilio Recabarren 2477, Iquique, Chile}

\begin{abstract}

This paper investigates the integrability properties of Einstein's theory of gravity in the context of accelerating Newman-Unti-Tamburino (NUT) spacetimes by utilizing Ernst's description of stationary and axially symmetric electrovacuum solutions. We employ Ehlers transformations, Lie point symmetries of the Einstein field equations, to efficiently endorse accelerating metrics with a nontrivial NUT charge. Under this context, we begin by rederiving the known C-metric NUT spacetime described by Chng, Mann, and Stelea in a straightforward manner, and in the new form of the solution introduced by Podolsk\'y and Vr\'atn\'y. Next, we construct for the first time an accelerating NUT black hole dressed with a conformally coupled scalar field. These solutions belong to the general class of type I spacetimes and, therefore, cannot be obtained from any limit of the Pleban\'ski-Demia\'nski family whatsoever and their integration needs to be carried out independently. 
Including Maxwell fields is certainly permitted, however, the use of Ehlers transformations is subtle and requires further modifications. Ehlers transformations  not only partially rotate the mass parameter such that its magnetic component appears, but also rotate the corresponding gauge fields. Notwithstanding, the alignment of the electromagnetic potentials can be successfully performed via a duality transformation, hence providing a novel Reissner-Nordstr\"om-C-metric NUT black hole that correctly reproduces the Reissner-Nordstr\"om-C-metric and Reissner-Nordstr\"om-NUT configurations in the corresponding limiting cases. 
We describe the main geometric features of these solutions and discuss possible embeddings of our geometries in external electromagnetic and rotating backgrounds.
\end{abstract}


\maketitle

\section{Introduction}

In the understanding of the gravitational interaction exact solutions of Einstein field equations play a fundamental role. In many cases these solutions allow one to study, by means of analytic computations, several phenomena of classical and semiclassical gravity with significant astrophysical implications. As a matter of fact, the most iconic exact spacetimes are given by black holes, today widely accepted to be primordial for the formation of galaxies \cite{Cattaneo:2009ub,Fabian:2012xr} and the production of gravitational waves as a result of binary black hole mergers \cite{LIGOScientific:2016aoc}. In addition, their mechanic and thermodynamic features are essential in the semiclassical description of gravitational systems, providing a first glance on the interplay of gravity and quantum field theory.
It is fair to state that the Schwarzschild \cite{Schwarzschild:1916uq} and Kerr spacetimes \cite{Kerr:1963ud}, describing the gravitational field of a static and a rotating source respectively, are the most famous black hole spacetimes in general relativity (GR). Along with their charged extensions, the Reissner-Nordstr\"om \cite{Reissner,Nordstrom} and Kerr-Newman \cite{Newman:1965my} spacetimes, and the so-called C-metric \cite{Kinnersley:1969zz} and Taub-NUT geometries \cite{Taub:1950ez,Newman:1963yy}, known to represent accelerating black holes and twisted spacetimes with a gravitomagnetic interpretation respectively, they make part of the general class of spacetimes known as the Pleban\'ski-Demia\'nski family \cite{Plebanski:1976gy}. The Pleban\'ski-Demia\'nski spacetime is the most general type D solution of Einstein-Maxwell equations and it represents a pair of charged rotating black holes, endowed with Newman-Unti-Tamburino (NUT) charge, that accelerate from each other in opposite directions. Including the presence of a cosmological constant, this family depends on seven continuous parameters, which can be (although not straightforwardly) identified with the mass $m$, angular momentum $a$, NUT charge $l$, acceleration parameter $\alpha$, and electric $e$ and magnetic $g$ charges. In practice, subfamilies of black holes and their combinations need to be obtained from the general Pleban\'ski-Demia\'nski line element \cite{Plebanski:1976gy} by performing special limiting procedures, a process that differs from being direct and that, on occasions, clouds the interpretation of the metric parameters \cite{Stephani:2003tm}.
The Pleban\'ski-Demia\'nski geometry has been initially studied by several authors, especially the case  which represents a spinning C-metric, a pair of rotating black holes accelerating from each other in opposite directions \cite{Griffiths:2005se,Griffiths:2005mi,Griffiths:2005qp,Pravda:2002kj,Bicak:1999sa}. Along these lines, Hong and Teo \cite{Hong:2003gx} found a convenient set of coordinates in which the metric polynomials of the C-metric can be easily factorized. This  allowed one not only to rapidly identify the corresponding Killing horizons but also to perform a deeper analysis of the causal structure of this spacetime. 
These findings motivated Griffiths and Podolsk\'y \cite{Griffiths:2005se} to perform a similar construction for the Pleban\'ski-Demia\'nski metric, which on this new form\footnote{In this paper we focus on the case in which the cosmological constant vanishes. This is strictly related with the fact that the generating solution technique we shall use to construct our new geometries does not apply for $\Lambda\neq0$.} depends on the six independent parameters $m$, $a$, $\alpha$, $l$, $e$, and $g$ and a seventh parameter, the so-called twisted parameter $\omega$ that is ultimately related to both, angular momentum and NUT charge. In this form almost all limits on the Pleban\'ski-Demia\'nski metric are direct to take, from which highlight the spinning C-metric and the Kerr-NUT spacetimes and their charged generalizations. As the Pleban\'ski-Demia\'nski line element represents at the end of the day a sort of spinning C-metric with NUT charge, it seems natural to wonder what the nonrotational limit of this spacetime is. In this form of the line element, there is no direct obstruction for this limit to be performed, and actually a simple stationary metric is obtained. However, this metric is found to  not possess any conical singularities, as it is known, the sources of the acceleration, and, even more, the acceleration parameter $\alpha$ becomes a redundant parameter that can be removed by a suitable coordinate transformation. It was then conjectured that an accelerating NUT line element does not exist, at least in the wide class of the Pleban\'ski-Demia\'nski family. Notwithstanding, an accelerating NUT family of black holes was introduced by Chng, Mann, and Stelea \cite{Chng:2006gh} by using a precise generating solution technique based on the symmetries of the dimensionally reduced Lagrangian of GR obtained by a timelike compactification from dimension four to three. This solution, which in fact does represent a NUT black hole with a nontrivial acceleration, was proven to  not belong to the special type D family given by Pleban\'ski-Demia\'nski and to have general algebraic properties and, in fact, is of the general type I. As a consequence, this metric could have not been obtained by any limiting process acting on the Pleban\'ski-Demia\'nski solution. A more convenient form of the metric and a full geometric analysis of this novel solution was given by Podolsk\'y and Vr\'atn\'y \cite{Podolsky:2020xkf}, where Killing horizons, algebraic classification, conformal completion and causal structures were studied in detail. Recently, the same authors have found a refined set of coordinates to represent the Pleban\'ski-Demia\'nski metric, with and without a cosmological constant \cite{Podolsky:2021zwr,Podolsky:2022xxd},\footnote{An extension of the Pleban\'ski-Demia\'nski solutions with a cosmological constant has been constructed
in the framework of  metric-affine gravity   \cite{Bahamonde:2022meb}.} and have finally cleaned out the presence of the nonindependent twist parameter $\omega$. This represents the final form of the Pleban\'ski-Demia\'nski spacetime and is constructed in such a manner that all limiting cases are clearly obtained with no need for further changes of coordinates; it reads\footnote{We follow the same notation given by Podolsk\'y and Vr\'atn\'y \cite{Podolsky:2021zwr,Podolsky:2022xxd} and constrain ourselves to the case $\Lambda=0$.}
\begin{align}
d s^2=\frac{1}{\Omega^2}\left(  -\frac{Q}{\rho^2}\left[d t-\left(a \sin ^2 \theta+4 l \sin ^2 \frac{1}{2} \theta\right) d \varphi\right]^2+\frac{\rho^2}{Q} d r^2  +\frac{\rho^2}{P} d \theta^2+\frac{P}{\rho^2} \sin ^2 \theta\left[a d t-\left(r^2+(a+l)^2\right) d \varphi\right]^2\right),
\end{align}
where
\begin{align}
\Omega & =1-\frac{\alpha a}{a^2+l^2} r(l+a \cos \theta), \\
\rho^2 & =r^2+(l+a \cos \theta)^2, \\
P(\theta) & =\left(1-\frac{\alpha a}{a^2+l^2} r_{+}(l+a \cos \theta)\right)\left(1-\frac{\alpha a}{a^2+l^2} r_{-}(l+a \cos \theta)\right), \\
Q(r) & =\left(r-r_{+}\right)\left(r-r_{-}\right)\left(1+\alpha a \frac{a-l}{a^2+l^2} r\right)\left(1-\alpha a \frac{a+l}{a^2+l^2} r\right).
\end{align}
The roots of the metric polynomial $Q(r)$ define the two black hole Killing horizons
\begin{align}
& r_{+}:=m+\sqrt{m^2+l^2-a^2-e^2-g^2}, \\
& r_{-}:=m-\sqrt{m^2+l^2-a^2-e^2-g^2},
\end{align}
that are accompanied by two accelerating horizons located at
\begin{equation}
r_\alpha:= \pm \frac{1}{\alpha} \frac{a^2+l^2}{a (a \pm l)}.
\end{equation}
The $a \rightarrow 0$ limit is, therefore, completely transparent, showing the outsider behavior of the NUT C-metric black hole with respect to the Pleban\'ski-Demia\'nski family. Indeed, $\Omega$ and $P$ tend to one; no conical singularities emerge and, therefore, no acceleration whatsoever. The metric simply represents a charged NUT black hole.

In this work, we aim to rederive the Chng, Mann, and Stelea solution  \cite{Chng:2006gh} by making use of a different, but theoretically close, solution generating technique. The Ernst scheme \cite{Ernst:1967wx,Ernst:1967by} describes stationary and axially symmetric spacetimes in Einstein-Maxwell theory. It is known that Einstein-Maxwell theory written in terms of the Ernst potentials enjoys a set of Lie point symmetries known as Ehlers symmetries \cite{Stephani:2003tm}, from which the nontrivial, Harrison \cite{HARRISON} and Ehlers \cite{Ehlers:1957zz} transformations can be used to obtain novel solutions in electrovacuum. In their electric and magnetic versions,\footnote{Electric and magnetic versions differ by a double Wick rotation on the corresponding Levis-Weyl-Papapetrou (LWP) line element, and their names are given by the effect that, in particular, Harrison transformations exert on a given seed spacetime.} Harrison transformations electrify a given seed spacetime or embed the seed on an electromagnetic background, respectively \cite{Ernst:1976mzr,Vigano:2022hrg,Siahaan:2021ags,Ghezelbash:2021lcf}. On the other hand, Ehlers transformations are known to add NUT charge or to embed the seed on a rotating (swirling) background \cite{Astorino:2022aam}. We shall apply an Ehlers transformation on the standard C-metric, written in intuitive spherical coordinates, to find in a few steps the convenient form of the Chng, Mann, and Stelea solution \cite{Chng:2006gh} provided by Podolsk\'y and Vr\'atn\'y \cite{Podolsky:2020xkf}. 
Then we present, for the first time, an accelerating NUT black hole dressed by a conformally coupled scalar field. This requires a specific treatment in order for the Ernst scheme to be applicable \cite{Astorino:2013xc}. Several solutions of this kind have been previously reported \cite{Astorino:2012zm,Astorino:2013sfa,Astorino:2013xxa,Astorino:2014mda,Astorino:2022prj}; however, none of them are in the context of accelerating NUT spacetimes. In technical terms, this is due to the algebraically general nature of accelerating NUT metrics. A natural step forward is to endorse these spacetimes with electromagnetic charges. At first glance, this could be achieved by simply adding NUT charge via an Ehlers transformation acting onto a Reissner-Nordstr\"om C-metric. Nonetheless, it has been shown that in order to add NUT parameter on a given charged spacetime  an enhanced Ehlers transformation it is requiered that, along with the usual Ehlers map, includes an extra duality transformation that restores the alignment of the corresponding gauge fields \cite{Astorino:2019ljy}. As a matter of fact, Ehlers transformations rotate the mass parameter in such a manner that its magnetic component becomes nontrivial but rotating at the same time the electromagnetic potentials. 
Here, we construct the NUT extension of the Reissner-Nordstr\"om C-metric spacetime, showing how the corresponding line element correctly reproduces the NUT Reissner-Nordstr\"om limit with an aligned gauge potential if the naive Ehlers map used in the uncharged case is improved with a simple duality transformation.

This paper is structured as follows: In Sec. \ref{secii}, we shortly review the Ernst scheme for stationary and axially symmetric solutions in Einstein-Maxwell theory. We put particular emphasis in explaining how to produce new solutions by means of Harrison and Ehlers symmetry transformations. Section \ref{seciii} is devoted to rederive the Chng, Mann, and Stelea solution \textit{à la} Podolsk\'y and Vr\'atn\'y by making use of an electric Ehlers transformation. We continue with Sec. \ref{seciv} extending our findings to the case in which a conformally coupled scalar field is introduced in the matter sector. We explain how the Ernst scheme is extended to this theory and how to integrate such a solution. Section \ref{secv} is destined to the construction of NUT Reissner-Nordstr\"om C-metric configurations, where we explicitly show the line element, its corresponding nonaccelerating limit, and how the alignment of the gauge field can be retrieved by a duality transformation.
Finally, we conclude in Sec. \ref{secvi} with a discussion of our results and with a list of possible avenues that can be explored to continue studying these geometries.

\section{Ernst scheme}\label{secii}
The Ernst scheme \cite{Ernst:1967wx,Ernst:1967by} provides an elegant framework in which to study stationary and axially symmetric spacetimes within Einstein-Maxwell theory. It usefulness relies in the ability it has to disclose certain symmetries of the electrovacuum theory that allow the generation of nontrivial solutions starting from a known seed. In concrete terms, Einstein-Maxwell field equations for a general stationary and axisymmetric spacetime given by the  (LWP) metric and a stationary and axially symmetric Maxwell field
\begin{align}
d s^2 & =-f(d t-\omega d \varphi)^2+f^{-1}\left[\rho^2 d \varphi^2+e^{2 \gamma}\left(d \rho^2+d z^2\right)\right],  \label{ELWP}\\
A & =A_t d t+A_{\varphi} d \varphi,
\end{align}
can be cast in a completely equivalent form, the Ernst equations
\begin{subequations}
\label{ernsteq}
\begin{align}
& \left(\operatorname{Re} \mathcal{E}+|\Phi|^2\right) \nabla^2 \mathcal{E}=\nabla \mathcal{E} \cdot\left(\nabla \mathcal{E}+2 \Phi^* \nabla \Phi\right), \\
& \left(\operatorname{Re} \mathcal{E}+|\Phi|^2\right) \nabla^2 \Phi=\nabla \Phi \cdot\left(\nabla \mathcal{E}+2 \Phi^* \nabla \Phi\right),
\end{align}
\end{subequations}
where $\vec{\nabla}$ and the various vectorial quantities are understood in Euclidean space with cylindrical coordinates $(\rho, \varphi, z)$. The so-called Ernst potentials are defined as
\begin{equation}
\mathcal{E}=f-|\Phi|^2+i \chi, \quad \Phi=A_t+i \tilde{A}_{\varphi},
\end{equation}
where $\tilde{A}_{\varphi}$ and $\chi$, known as twisted potentials, are characterized via the differential equations
\begin{equation}
\hat{\varphi} \times \nabla \tilde{A}_{\varphi}=\rho^{-1} f\left(\nabla A_{\varphi}+\omega \nabla A_t\right) \label{twistaphi}
\end{equation}
and
\begin{equation}
\hat{\varphi} \times \nabla \chi=-\rho^{-1} f^2 \nabla \omega-2 \hat{\varphi} \times \operatorname{Im}\left(\Phi^* \nabla \Phi\right). \label{eqchi}
\end{equation}
It is important to remark that, due to the integrability features of the Einstein-Maxwell system, the function $\gamma(\rho, z)$ decouples, and it is uniquely determined by the functions $f(\rho, z)$ and $\omega(\rho, z)$.
This form of the electrovacuum field equations can be proven to enjoy a set of Lie point symmetries, known as Ehlers symmetries \cite{Stephani:2003tm}, given by
\begin{subequations}
\label{ernst-group}
\begin{align}
\label{gauge1}
\ernst & = |\lambda|^2 \ernst_0 \,, \qquad\qquad\qquad\quad\;\;
\Phi = \lambda \Phi_0 \,, \\
\label{gauge2}
\ernst & = \ernst_0 + ib \,, \qquad\qquad\qquad\quad\,
\Phi = \Phi_0 \,, \\
\label{ehlers}
\ernst & = \frac{\ernst_0}{1 + i\jmath\ernst_0} \,, \qquad\qquad\qquad\;\,
\Phi = \frac{\Phi_0}{1 + i\jmath\ernst_0} \,,  \\
\label{gauge3}
\ernst & = \ernst_0 - 2\beta^*\Phi_0 - |\beta|^2 \,, \quad\quad\hspace{0.1cm}
\Phi = \Phi_0 + \beta \,, \\
\label{harrison}
\ernst & = \frac{\ernst_0}{1 - 2\alpha^*\Phi_0 - |\alpha|^2\ernst_0} \,, \quad \hspace{0.2cm}
\Phi = \frac{\alpha\ernst_0 + \Phi_0}{1 - 2\alpha^*\Phi_0 - |\alpha|^2\ernst_0} \,, 
\end{align}
\end{subequations}
where $\alpha, \beta$, and $\lambda$ are complex parameters, while $b$ and $\jmath$ are real. After reducing the pure gauge transformations we are left with \eqref{ehlers} and \eqref{harrison}, the so-called Ehlers \cite{Ehlers:1957zz} and Harrison \cite{HARRISON}  transformations. Because of their Lie point symmetry nature, the action of the transformations leaves the Ernst equations unchanged while at the same time producing new nonequivalent geometries. Notice that LWP metric \eqref{ELWP} is not the unique choice for a general stationary and axisymmetric line element. As a matter of fact, we can act on it with a discrete double-Wick rotation producing the nonequivalent configuration
\begin{align}
d s^2 & =f(d \varphi-\omega d t)^2+f^{-1}\left[e^{2 \gamma}\left(d \rho^2+d z^2\right)-\rho^2 d t^2\right] \label{MLWP}, \\
A & =A_t d t+A_{\varphi} d \varphi,
\end{align}
that gives rise to the very same Ernst equations \eqref{ernsteq}. Here, $\Phi=A_{\varphi}+i \tilde{A}_t$ and
\begin{equation}
\hat{\varphi} \times \nabla \tilde{A}_t=\rho^{-1} f\left(\nabla A_t+\omega \nabla A_{\varphi}\right).
\end{equation}
This ansatz is usually referred to as the magnetic LWP ansatz. Then, the effect of Harrison and Ehlers transformations on a given seed spacetime crucially depends on which ansatz they act \cite{Vigano:2022hrg}, i.e., \eqref{ELWP} or \eqref{MLWP}. The Harrison map is known to electrify a given seed when acting on \eqref{ELWP}, while it embeds a seed spacetime of the form \eqref{MLWP} in an electromagnetic universe  \cite{Ernst:1976mzr}. On the other hand, Ehlers transformations add a NUT parameter to a given seed when acting on \eqref{ELWP} while embedding the initial spacetime in a rotating (swirling) background when acting on \eqref{MLWP} \cite{Astorino:2022aam}. 

For our purposes, it is useful to consider the enhancement in which a minimally coupled scalar field is introduced in the matter sector. It was proven in Ref. \cite{Astorino:2013xc} that the Ernst scheme survives the addition of such a field; in fact, the reduced action principle in terms of the Ernst potentials reveals that in the Einstein-Maxwell-scalar case the set of Ehlers symmetries \eqref{ernst-group} needs to be complemented by the trivial transformation $\Psi \rightarrow \Psi^{\prime}=\Psi$, while Ernst equations are complemented with the corresponding Klein-Gordon equation for $\Psi$. This ensures the feasibility of Ehlers transformations as an effective solution generating technique when dealing with black holes with scalar hair. Most of spacetimes with scalar hair are easily found in the so-called Einstein-Maxwell-conformal-scalar theory; this means in the Jordan frame in which the scalar field is conformally coupled to the curvature scalar \cite{Bocharova:1970skc,Bekenstein:1975ts,Charmousis:2009cm,Anabalon:2009qt,Bhattacharya:2013hvm,Bardoux:2013swa,Cisterna:2021xxq}. Both theories, with minimally and conformally coupled scalar fields, are related by means of the conformal map known as Bekenstein transformations \cite{Bekenstein:1974sf}. These transformations provide a vehicle to transport all the machinery of the Ernst scheme to theories with conformally coupled scalars. It is then enough to consider a seed spacetime in the Jordan frame, transform it to the minimally coupled frame (in which Ehlers transformations can be applied) and move back to the conformally couple theory to obtain new solutions in the Einstein-Maxwell-conformal-scalar sector.

With these tools at hand, in the next sections we will rederive the Chng, Mann, and Stelea solution \cite{Chng:2006gh}, apply these results to the case in which a conformally coupled scalar field is considered in the matter sector, and explore the construction of charged accelerating NUT geometries.

\section{Dressing the C-metric with NUT}\label{seciii}

The aim of this section is to construct the accelerating NUT black hole of Chng, Mann, and Stelea \cite{Chng:2006gh}, written in the coordinates provided by Podolsk\'y and Vr\'atn\'y \cite{Podolsky:2020xkf}. The steps are very clear. We take the known C-metric in spherical-like coordinates, and we dress it with a NUT charge via an electric Ehlers transformation \cite{Ehlers:1957zz}. Thus, we start from
\begin{equation}
d s^2=\frac{1}{\Omega(R, \theta)^2}\left[-Q(R) d t^2+\frac{d R^2}{Q(R)}+\frac{R^2 d \theta^2}{P(\theta)}+R^2 P(\theta) \sin ^2 \theta d \varphi^2\right], \label{cmetricseed}
\end{equation}
where
\begin{align}
\Omega(R, \theta) & =1+A R \cos \theta, \\
Q(R) & =\left(1-A^2 R^2\right)\left(1-\frac{2 M}{R}\right), \\
P(\theta) & =1+2 A M \cos \theta .
\end{align}
Here, $A$ and $M$ stand for the acceleration parameter and the mass of the black hole, respectively \cite{Griffiths:2009dfa}. The standard use of the Ernst scheme requires one to compare metric \eqref{cmetricseed} with the electric LWP ansatz \eqref{ELWP}, from which it is direct to recognize the seed functions
\begin{subequations}
\label{seedfunctions}
\begin{align}
f_0 & =\frac{Q}{\Omega^2}, \\
\rho & =\frac{\sqrt{P Q} R \sin \theta}{\Omega^2}, \\
\omega_0 & =0 .
\end{align}
\end{subequations}
This is done in order to write the seed Ernst potentials. Note that seed quantities are denoted by the subindex 0. Therefore, the seed potentials are given by
\begin{equation}
\mathcal{E}_0=\frac{Q}{\Omega^2}, \quad \Phi_0=0,
\end{equation}
where indeed the potential $\Phi_0$ vanishes and the seed is uncharged. In addition, $\mathcal{E}_0$ is real, corresponding with the static character of the seed metric. With these at hand, we apply the Ehlers transformation \footnote{Aiming to maintain the notation given by Podolsk\'y and Vr\'atn\'y, we change the real parameter $\jmath$ to $c$.} \eqref{ehlers} and obtain the transformed Ernst potential (free of any subindex)
\begin{equation}
\mathcal{E}=\frac{\mathcal{E}_0}{1+i c \mathcal{E}_0}=\frac{Q / \Omega^2}{\Lambda},
\end{equation}
where we have defined the quantity
\begin{equation}
\Lambda(R, \theta)=1+i c \frac{Q}{\Omega^2}.
\end{equation}
The electromagnetic potential remains null: $\Phi=0$. To read the transformed metric components, we make use of the definition of the Ernst potential, which now is composed by real and imaginary parts. From the real part, we identify
\begin{equation}
f=\operatorname{Re}(\mathcal{E})=\frac{Q / \Omega^2}{|\Lambda|^2}=\frac{f_0}{|\Lambda|^2},
\end{equation}
where $|\Lambda|^2=\Lambda \bar{\Lambda}$. The imaginary part of $\mathcal{E}$,
\begin{equation}
\chi=\operatorname{Im}(\mathcal{E})=-c \frac{Q^2 / \Omega^4}{|\Lambda|^2},
\end{equation}
provides us, via Eq. \eqref{eqchi},\footnote{We use the orthonormal frame defined by the ordered triad $\left(\vec{e}_R, \vec{e}_{\varphi}, \vec{e}_\theta\right)$, so then $\vec{e}_{\varphi} \times \vec{e}_R=-\vec{e}_\theta$ and $\vec{e}_{\varphi} \times \vec{e}_\theta=\vec{e}_R$. The gradient operator is then defined as $\vec{\nabla}\propto\sqrt{Q}R\frac{\partial}{\partial R}\vec{e}_R+\sqrt{P}\frac{\partial}{\partial \theta}\vec{e}_\theta+\frac{1}{\sqrt{P}\sin\theta}\frac{\partial}{\partial \varphi}\vec{e}_\varphi$.} with the form of the rotating function $\omega$:
\begin{equation}
\omega(R, \theta)=2 c\left(2 M \cos \theta+\frac{A P(\theta) R^2 \sin ^2 \theta}{\Omega(R, \theta)^2}\right).
\end{equation}
Ehlers transformations always endorse an imaginary part on the transformed Ernst potential. Therefore, it always implies the appearance of some sort of rotation [via \eqref{eqchi}], NUT charge in the electric case, or swirling rotation in the magnetic one \cite{Vigano:2022hrg}. Recalling that Ehlers transformations do not affect the metric function $\gamma, \gamma=\gamma_0$, it is possible to verify that
\begin{equation}
e^{2\gamma_0}\left(d \rho^2+d z^2\right)=e^{2\gamma}\left(d \rho^2+d z^2\right)=\left(\frac{d R^2}{Q}+\frac{R^2}{P} d \theta^2\right) f_0,
\end{equation}
a result that, along with the definitions for $f$ and $\omega$, can be plugged into the LWP ansatz \eqref{ELWP} to deliver the transformed metric
\begin{align}
d s^2= & \frac{1}{\Omega(R, \theta)^2}\left[-\frac{Q(R)}{|\Lambda(R, \theta)|^2}\left[d t-2 c\left(2 M \cos \theta+\frac{A P(\theta) R^2}{\Omega(R, \theta)^2} \sin ^2 \theta\right) d \varphi\right]^2\right. \nonumber\\
& \left.+|\Lambda(R, \theta)|^2\left(R^2 P(\theta) \sin ^2 \theta d \varphi^2+\frac{d R^2}{Q(R)}+R^2 \frac{d \theta^2}{P(\theta)}\right)\right] .
\end{align}
At this stage, the construction is completed, and it  remains only to establish the connection between the real parameter $c$ and the NUT charge $l$. For this to be done, we define a new mass parameter via the relation
\begin{equation}
m=\sqrt{M^2-l^2},
\end{equation}
which together with the definitions
\begin{align}
& r_{+} \equiv m+\sqrt{m^2+l^2}, \\
& r_{-} \equiv m-\sqrt{m^2+l^2}
\end{align}
allows one to find a proper change of coordinates to bring down the metric in an explicit NUT form. Keeping in mind that
\begin{align}
r_{+}+r_{-} & =2 m, \\
r_{+}-r_{-} & =2 \sqrt{m^2+l^2}, \\
r_{+}\left(r_{+}-r_{-}\right) & =r_{+}^2+l^2,
\end{align}
we perform the change of coordinates $R \rightarrow\left(r-r_{-}\right)$ and $t \rightarrow \frac{\left(r_{+}-r_{-}\right)}{r_{+}} \tau$ together with the reparametrization $c \rightarrow l / r_{+}$. The metric terms change as
\begin{align}
&-\frac{Q(R)}{|\Lambda(R, \theta)|^2} {\left[d t-2 c\left(2 M \cos \theta+\frac{A P(\theta) R^2}{\Omega(R, \theta)^2} \sin ^2 \theta\right) d \varphi\right]^2 }  \rightarrow-\frac{r_{+}-r_{-}}{r_{+}} \frac{\mathcal{Q}(r)}{\mathcal{R}^2(r, \theta)}\left[d \tau-2 l\left(\cos \theta+A \mathcal{T}(r, \theta) \sin ^2 \theta\right) d \varphi\right]^2, \\
&|\Lambda(R, \theta)|^2 R^2 \rightarrow \frac{r_{+}-r_{-}}{r_{+}} \mathcal{R}^2(r, \theta), \quad \frac{|\Lambda(R, \theta)|^2}{Q(R)} \rightarrow \frac{r_{+}-r_{-}}{r_{+}} \frac{\mathcal{R}^2(r, \theta)}{\mathcal{Q}(r)},
\end{align}
where we have defined the new metric functions
\begin{align}
\Omega(r, \theta) & =1+A\left(r-r_{-}\right) \cos \theta, \\
\mathcal{P}(\theta) & =1+A\left(r_{+}-r_{-}\right) \cos \theta, \\
\mathcal{Q}(r) & =\left(1-A^2\left(r-r_{-}\right)^2\right)\left(r-r_{-}\right)\left(r-r_{+}\right), \\
\mathcal{T}(r, \theta) & =\frac{P(\theta)\left(r-r_{-}\right)^2}{\left(r_{+}-r_{-}\right) \Omega(r, \theta)^2}, \\
\mathcal{R}^2(r, \theta) & =\frac{1}{r_{+}^2+l^2}\left[r_{+}^2\left(r-r_{-}\right)^2+l^2 \frac{\left[1-A^2\left(r-r_{-}\right)^2\right]^2}{\Omega(r, \theta)^4}\left(r-r_{+}\right)^2\right] .
\end{align}
Up to a conformal factor that can be disregarded by a conformal rescaling of the line element, we find the final form of the spacetime to be
\begin{align}
d s^2= & \frac{1}{\Omega(r, \theta)^2}\left[-\frac{\mathcal{Q}(r)}{\mathcal{R}^2(r, \theta)}\left[d \tau-2 l\left(\cos \theta+A \mathcal{T}(r, \theta) \sin ^2 \theta\right) d \varphi\right]^2+\mathcal{R}^2(r, \theta)\left(\mathcal{P}(\theta) \sin ^2 \theta d \varphi^2+\frac{d r^2}{\mathcal{Q}(r)}+\frac{d \theta^2}{\mathcal{P}(\theta)}\right)\right]. \label{lineelementmann}
\end{align}
This precisely coincides with the accelerating NUT spacetime presented by Podolský and Vrátný \cite{Podolsky:2020xkf}. The corresponding limits are clean, either $A \rightarrow 0$ or $l \rightarrow 0$; thus, we recover the NUT or C-metric black holes with no further changes of coordinates. We do not intend to provide a detailed description of the geometric features of this solution, as it has been already done in Ref. \cite{Podolsky:2020xkf}.

\section{Accelerating NUT black holes with a conformally coupled scalar dress}\label{seciv}

The Ernst solution generating technique is, in principle, intrinsic to four-dimensional Einstein-Maxwell theory; then it is expected that Ehlers transformations  apply only in the context of black holes with scalar hair. However, a few counterexamples have been found. A particular symmetry of the Einstein-dilaton-Maxwell theory has been identified, of which the effect is to embed a given seed on an external magnetic field \cite{Dowker:1993bt}. Moreover, this system has been shown to exhibit internal symmetries allowing for a deeper exploration of its black hole spectrum \cite{Galtsov:1994pd,Galtsov:1994sjr}. On the other hand, in higher dimensions, a magnetizing transformation has been also identified, providing Melvin solutions in arbitrary dimensions \cite{Ortaggio:2004kr}. A particularly appealing extension of the Ernst scheme is the one in which a minimally coupled scalar field is included in the matter sector \cite{Astorino:2013xc}. As explained before, Harrison and Ehlers symmetries remain as valid Lie point transformations of this theory and, therefore, open the road for the exploration of the integrability properties of Einstein-Maxwell-scalar theory in a wide variety of cases. Making use of Bekenstein transformations \cite{Bekenstein:1974sf}, all these new solutions can be mapped to the Einstein-Maxwell-conformal-scalar scenario, in which they often represent black hole spacetimes. In this section, we apply all this machinery to construct an accelerating NUT black hole with a conformally coupled scalar dress. Black holes with a conformally coupled scalar field have been deeply investigated in many contexts, but in the case where acceleration and NUT charge are combined in the absence of rotation. This is, again, due to the fact that when doing so we depart from the Pleban\'ski-Demia\'nski family. The strategy follows from the previous section, now augmented with the use of the Bekenstein map. 

\begin{figure}[H]
\begin{center}
\includegraphics[scale=1]{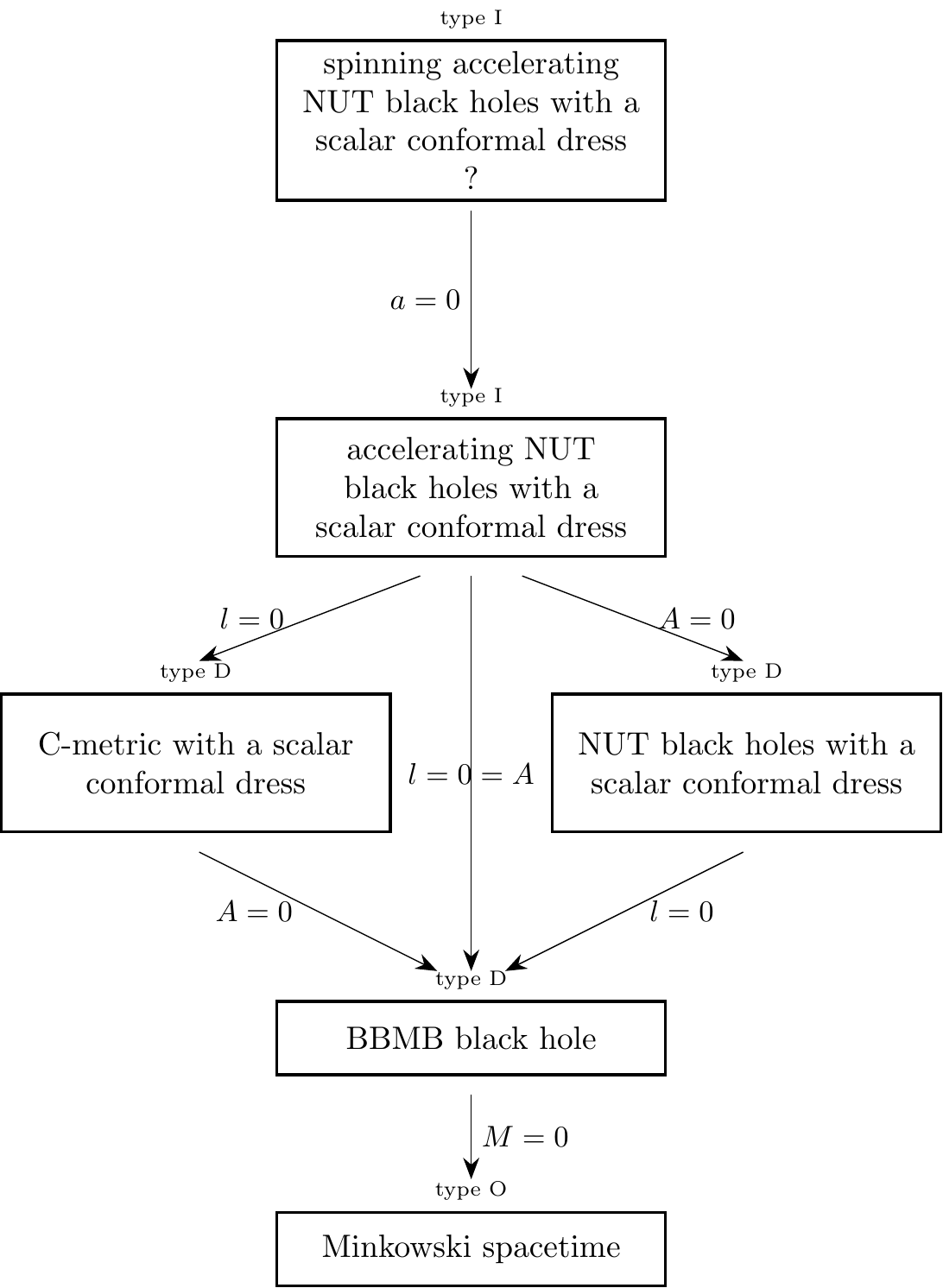}
\end{center}
\caption{This diagram represents a hierarchy of solutions with a scalar conformal dress that continuously connect with the solution described in this section. On top, there is a spacetime that includes rotation, acceleration, and NUT charge and that might be obtained by adding rotation on our spacetime \eqref{solutionnewscalar}. As discussed in our conclusions, it is expected to be of type I and, therefore has not been yet found. The corresponding limits $l=0$ and/or $A=0$ bring us to the well-known type D solutions with scalar conformal dress.}
\label{fig1}
\end{figure}

Thus, we start by considering a C-metric spacetime in the context of a conformally coupled scalar theory. We proceed by translating the solution to the Einstein frame by means of an inverse Bekenstein map and by applying an electric Ehlers transformation to add the corresponding NUT charge. The final form of the solution is obtained by moving back to the Jordan frame.
Our guide action is 
\begin{equation}
I=\frac{1}{2 \kappa} \int d^4 x \sqrt{-g}\left[R-\kappa\left(g^{\mu \nu} \partial_\mu \phi \partial_\nu \phi+\frac{1}{6} R \phi^2\right)\right],
\end{equation}
for which C-metric black holes with conformal scalar hair have been identified in Refs. \cite{Charmousis:2009cm,Anabalon:2009qt}. In terms of the acceleration conformal factor, the metric and scalar are written, respectively, as
\begin{align}
d s^2 & =\frac{1}{\Omega(R, \theta)^2}\left[-\frac{Q(R)}{R^2} d t^2+\frac{R^2}{Q(R)} d R^2+\frac{R^2}{P(\theta)} d \theta^2+P(\theta) R^2 \sin ^2 \theta d \varphi^2\right], \\
\phi(R, \theta) & =\sqrt{\frac{6}{\kappa}} \frac{k \Omega(R, \theta)}{R+k(\Omega(R, \theta)-2)}, \label{cmetricscalar}
\end{align}
where
\begin{subequations}
\label{cmetricfunctionR}
\begin{align}
\Omega(R, \theta) & =1+A R \cos \theta, \\
Q(R) & =\left(1-A^2 R^2\right)(R-M)\left(R-\frac{M}{1+2 A M}\right), \\
P(\theta) & =(1+A M \cos \theta)\left(1+\frac{A M}{1+2 A M} \cos \theta\right), \\
k & =\frac{M}{1+A M} .
\end{align}
\end{subequations}
In order to proceed, we move to the Einstein frame
\begin{equation}
I=\frac{1}{2 \kappa} \int d^4 x \sqrt{-\tilde{g}}\left[\tilde{R}-\kappa \tilde{g}^{\mu \nu} \partial_\mu \Psi \partial_\nu \Psi\right],
\end{equation}
where the metric and scalar read, respectively,
\begin{align}
d s^2 & =\frac{\Omega_B(R, \theta)}{\Omega(R, \theta)^2}\left[-\frac{Q(R)}{R^2} d t^2+\frac{R^2}{Q(R)} d R^2+\frac{R^2}{P(\theta)} d \theta^2+P(\theta) R^2 \sin ^2 \theta d \varphi^2\right], \\
\Psi(R, \theta) & =\sqrt{\frac{6}{\kappa}} \tanh ^{-1}\left(\sqrt{\frac{\kappa}{6}} \phi(R, \theta)\right) .
\end{align}
Here, the Bekenstein factor is given by
\begin{equation}
\Omega_B(R, \theta)=1-\frac{\kappa}{6} \phi(R, \theta)^2.
\end{equation}
Now, we apply an electric Ehlers transformation in order to add the NUT charge. We do not repeat every step, as they are identical to the previous case, no electromagnetic charges are here included, and the identification of the seed and transformed Ernst potentials follows in full analogy with those of the previous section.
Thus, we get 
\begin{equation}
d s^2=\frac{\Omega_B(R, \theta)}{\Omega(R, \theta)^2}\left[-\frac{Q(R)}{|\Lambda(R, \theta)|^2 R^2}(d t-\omega(R, \theta) d \varphi)^2+|\Lambda(R, \theta)|^2 R^2\left(P(\theta) \sin ^2 \theta d \varphi^2+\frac{d \theta^2}{P(\theta)}+\frac{d R^2}{Q(R)}\right)\right],
\end{equation}
with
\begin{equation}
|\Lambda(R, \theta)|^2=1+c^2 \frac{\Omega_B(R, \theta)^2 Q(R)^2}{\Omega(R, \theta)^4 R^4}, \quad \omega(R, \theta)=2 c(2 M \cos \theta-A\mathcal{T}(R, \theta)),
\end{equation}
and where the function $\mathcal{T}(R, \theta)$ is given by
\begin{equation}
\begin{aligned}
\mathcal{T}(R,\theta)&=\frac{1}{(1+AM)^2(1+2AM)\Omega(R,\theta)^2(R+k(\Omega(R,\theta)-2))^2}\left[2A^3M^3R^4(1+2AM)\cos^5\theta\right.\\
&\quad\left.+A^2M^2R^2(8A^2M^2R^2-6AM^2R+14AMR^2+6M^2-6MR+5R^2)\cos^4\theta\right.\\
&\quad\left.+2AMR^2(2+AM)(2A^2M^2R^2-2AM^2R+3AMR^2+2M^2-2MR+R^2)\cos^3\theta\right.\\
&\quad \left.+\left(12A^3M^4R^3-2A^4M^4R^4-6A^3M^3R^4-14A^2M^4R^2+12A^2M^3R^3+A^2M^2R^4\right.\right.\\
&\quad\left.\left.+4AM^4R-4AM^3R^2-2AM^2R^3+4AMR^4-2M^4+4M^3R-2MR^3+R^4\right)\right.\cos^2\theta\\
&\quad\left.-2M\left(A^4M^3R^4-5A^3M^3R^3+5A^3M^2R^4+4A^2M^3R^2-10A^2M^2R^3+6A^2MR^4\right.\right.\\
&\quad\left.\left.+2AM^3R+3AM^2R^2-5AMR^3+2AR^4-M^3+2M^2R-MR^2\right)\cos\theta\right.\\
&\quad\left.-2A^3M^3R^4+3A^2M^4R^2+4A^2M^3R^3-5A^2M^2R^4-6AM^4R+2AM^3R^2+6AM^2R^3\right.\\
&\quad\left.-4AMR^4+3M^4-6M^3R+M^2R^2+2MR^3-R^4\right].
\end{aligned}
\end{equation}
Contrary to the previous construction, the function $\chi$ is more involved, and usually solving \eqref{eqchi} becomes cumbersome. To circumvent these technicalities, we have solved $\omega$ from the different, but equivalent expression
\begin{equation}
\vec{\nabla} \omega=|\Lambda|^2 \vec{\nabla} \omega_0-i \hat{\varphi} \times \frac{\rho}{f_0}\left(\bar{\Lambda} \vec{\nabla} \Lambda-\Lambda \vec{\nabla} \bar{\Lambda}\right).
\end{equation}
Hence, back to the conformal frame, the solution takes the form
\begin{equation}
d s^2=\frac{1}{\Omega(R, \theta)^2}\left[-\frac{Q(R)}{|\Lambda(R, \theta)|^2 R^2}(d t-\omega(R, \theta) d \varphi)^2+|\Lambda(R, \theta)|^2 R^2\left(P(\theta) \sin ^2 \theta d \varphi^2+\frac{d \theta^2}{P(\theta)}+\frac{d R^2}{Q(R)}\right)\right],
\end{equation}
where the scalar field and metric functions are given by \eqref{cmetricscalar} and \eqref{cmetricfunctionR}. As a final step, it  remains only to make up the metric in such a way that the NUT parameter naturally emerges. This proceeds in analogy with the case of accelerating NUT in GR, yielding
\begin{subequations}
\label{solutionnewscalar}
\begin{align}
&d s^2=\frac{1}{\bar{\Omega}(r, \theta)^2}\left[-\frac{\bar{\mathcal{Q}}(r)}{\overline{\mathcal{R}}^2(r, \theta)}[d \tau-2 l(\cos \theta-A\overline{\mathcal{T}}(r, \theta)) d \varphi]^2+\overline{\mathcal{R}}^2(r, \theta)\left(\bar{\mathcal{P}}(\theta) \sin ^2 \theta d \varphi^2+\frac{d \theta^2}{\bar{\mathcal{P}}(\theta)}+\frac{d r^2}{\bar{\mathcal{Q}}(r)}\right)\right], \\
&\bar{\phi}(r, \theta)=\sqrt{\frac{6}{\kappa}} \frac{k \bar{\Omega}(r, \theta)}{r-r_{-}+k(\bar{\Omega}(r, \theta)-2)},
\end{align}
\end{subequations}
where
\begin{subequations}
\label{nutcmetricconformally}
\begin{align}
& \bar{\Omega}(r, \theta)=1+A\left(r-r_{-}\right) \cos \theta, \\
& \bar{\mathcal{Q}}(r)=\left(1-A^2\left(r-r_{-}\right)^2\right)\left(r-r_{-}-\frac{r_{+}-r_{-}}{2}\right)\left(r-r_{-}-\frac{r_{+}-r_{-}}{2\left(1+A\left(r_{+}-r_{-}\right)\right)}\right), \\
& \bar{\mathcal{P}}(\theta)=\left(1+\frac{A\left(r_{+}-r_{-}\right)}{2} \cos \theta\right)\left(1+\frac{A\left(r_{+}-r_{-}\right)}{2\left(1+A\left(r_{+}-r_{-}\right)\right)} \cos \theta\right), \\
& \overline{\mathcal{T}}(r, \theta)=\frac{\mathcal{T}(r, \theta)}{r_{+}-r_{-}}, \\
& \overline{\mathcal{R}}^2(r, \theta)=\frac{1}{r_{+}^2+l^2}\left[r_{+}^2\left(r-r_{-}\right)^2+l^2 \frac{\bar{\Omega}_B(r, \theta)^2}{\bar{\Omega}(r, \theta)^4\left(r-r_{-}\right)^2} \bar{\mathcal{Q}}(r)^2\right] \text {. } 
&
\end{align}
\end{subequations}
Let us now briefly analyze the main features of this solution.\footnote{A detailed study of this geometry will be given in Ref.\cite{barrientosetall}.} It is instructive to start by taking the $l=0$ and $A=0$ limits. The former case, indeed, allows us to recover the original seed, namely, the accelerating solution with a conformally coupled scalar dress \cite{Charmousis:2009cm,Anabalon:2009qt}. Recalling that in the limit $l=0$ we have $r_+=2m$ and $r_-=0$, we can obtain  
\begin{subequations}
\begin{align}
&\bar{\Omega}(r,\theta)=1+Ar\cos\theta,\\
&\bar{P}(\theta)= (1+2Am\cos\theta)\left(1+\frac{Am\cos\theta}{1+2Am}\right),\\
&\bar{Q}(r)= (1-A^2r^2)(r-m)\left(r-\frac{m}{1+2Am}\right),\\
&\bar{\mathcal{R}}(r,\theta)=r^2,
\end{align}
\end{subequations}
on which we can replace $m$ by $M$ and $r$ by $R$, providing the seed metric configuration 
\begin{align}
&ds^2=\frac{1}{(1+AR\cos\theta)^2}\left[-\frac{\bar{Q}(R)}{R^2}dt^2+\frac{R^2dR^2}{\bar{Q}(R)}+\frac{d\theta^2}{\bar{P}(\theta)}+\bar{P}(\theta)r^2\sin^2\theta d\varphi\right],\\
&\phi(R,\theta)=\sqrt{\frac{6}{\kappa}}\frac{k(1+AR\cos\theta)}{R-k(1-AR\cos\theta)}.
\end{align}
More informative in terms of the NUT nature of our solution is the nonaccelerating limit. Taking $A\rightarrow0$ and recalling that $r_++r_-=2m$, we get 
\begin{subequations}
\begin{align}
&\bar{\Omega}(r,\theta)=1,\\
&\bar{P}(\theta)= 1,\\
&\bar{Q}(r)= (r-m)^2,\\
&\bar{\mathcal{R}}(r,\theta)=r^2+l^2,
\end{align}
\end{subequations}
that results into the NUT black hole with scalar conformal dress presented in Refs. \cite{Bhattacharya:2013hvm,Bardoux:2013swa}:
\begin{align}
&ds^2=-\frac{(r-m)^2}{(r^2+l^2)}(dt-2l\cos\theta d\varphi)^2+\frac{(r^2+l^2)}{(r-m)^2}dr^2+(r^2+l^2)(d\theta^2+\sin^2\theta d\varphi^2),\\
&\phi(r)=\sqrt{\frac{6}{\kappa}}\frac{\sqrt{m^2+l^2}}{r-m}.
\end{align}
This ensures that our NUT charge has been properly added into the original C-metric seed. As a matter of fact, both solutions, the seed and the NUT conformally coupled configuration, retrieve the well-known Bocharova-Bronnikov-Melnikov-Bekenstein
(BBMB) black hole \cite{Bocharova:1970skc,Bekenstein:1975ts} in the corresponding $A=0$ and $l=0$ limits, a solution that finally lands on the Minkowski spacetime in the vanishing mass case. As a matter of fact, the BBMB solution is not connected with the Schwarzschild black hole; there is no continuous limit that renders the spacetime free from the scalar field profile while keeping the mass. This hierarchy of scalar dressed solutions is depicted in \autoref{fig1}. 
The horizon structure of our accelerating NUT solution \eqref{solutionnewscalar} is easily obtainable due to the factorizable form of $\bar{Q}(r)$. Four Killing horizons are identified: 
\begin{subequations}
\label{killingscalar}
\begin{align}
&r_{a}^-=r_- - \frac{1}{A}=m-\sqrt{m^2+l^2}-\frac{1}{A},\\
&r_{a}^+=r_- + \frac{1}{A}=m-\sqrt{m^2+l^2}+\frac{1}{A},\\
&r_{H}^-=m-\frac{2A(m^2+l^2)}{1+2A\sqrt{m^2+l^2}},\\
&r_{H}^+=m,
\end{align}
\end{subequations}
which are given in two pairs, two acceleration horizons and two black hole horizons. Their appearance on the causal structure of \eqref{solutionnewscalar} is directly dependent on the range of the radial coordinate $r$. A brute-force inspection of the Kretschmann curvature scalar reveals, through a particularly lengthly and ugly expression, that the spacetime is free of curvature singularities. As a matter of fact, the initial singularity at $r=0$ of the seed configuration is completely softened by the addition of the NUT parameter, rendering our metric fully regular. In consequence, the four Killing horizons \eqref{killingscalar} are present in our geometry. It is direct to observe that $r_{a}^+>r_{a}^-$ and $r_{H}^+>r_{H}^-$; however, to completely determine the relative position of all Killing horizons with respect to themselves, an extra assumption is needed. A physically plausible condition is to consider the slowly acceleration regime, which, in turn, restricts the parameters $A$, $m$, and $l$ according with
\begin{equation}
A<\frac{1}{\sqrt{m^2+l^2}}=\frac{1}{M}. 
\end{equation} 
Hence, the horizons satisfy $r_{a}^-<r_{H}^-<r_{H}^+<r_{a}^+$. It is interesting to observe that, contrary to what occurs in the hairless case, here $r_{H}^-$ is not necessarily negative, and a causal structure  might arise  with three positive Killing horizons $r_{H}^-$, $r_{H}^+$, and $r_{a}^+$, namely, one acceleration horizon and a black hole and internal horizon. 

On the other hand, it is important to establish a possible pole of the scalar field profile ($r_0$), a pole that could render the scalar field profile singular on the outer domain of communications. In the case of \eqref{solutionnewscalar}, the $\theta$ dependence in the scalar field profile transforms such a locus in a region. In fact, it lies at 
\begin{equation}
r_0(\theta)=m-\frac{A(m^2+l^2)(\cos\theta+1)}{A\sqrt{m^2+l^2}(\cos\theta+1)+1}.
\end{equation}
We observe that for any value $\theta\neq\pi$ such a divergence of the scalar field profile always lies behind the event horizon $r_{H}^+$. However,  the conflictive locus  remains located at the south pole. This is an awkwardness that has been inherited from the seed ancestor and that can be removed by applying a conformal transformation to a very specific frame \cite{barrientosetall}. 

Accelerating and NUT spacetimes are, from the scratch, geometrically intricate solutions. It is known that the acceleration in the former case is due to the presence of conical defects, defects that are better represented by either cosmic strings or struts that pull or push the black hole to infinity, causing its acceleration. On the other hand, the NUT spacetime possesses the so-called Misner string topological defect, a sort of topological defect usually regarded as a semi-infinite rod with spin \cite{Bonnor:1969ala}. Solution \eqref{solutionnewscalar}, as it has been studied already in great detail in Ref. \cite{Podolsky:2020xkf} for the bold case,  also contains both type of topological structures. Its full analysis, along with a proper study of the algebraic classification of the solution (that here we have anticipated to be of the algebraically general type I), will be presented in Ref. \cite{barrientosetall}. 

\section{Reissner-Nordstr\"om accelerating NUT black hole}\label{secv}

\begin{figure}[H]
\begin{center}
\includegraphics[scale=1]{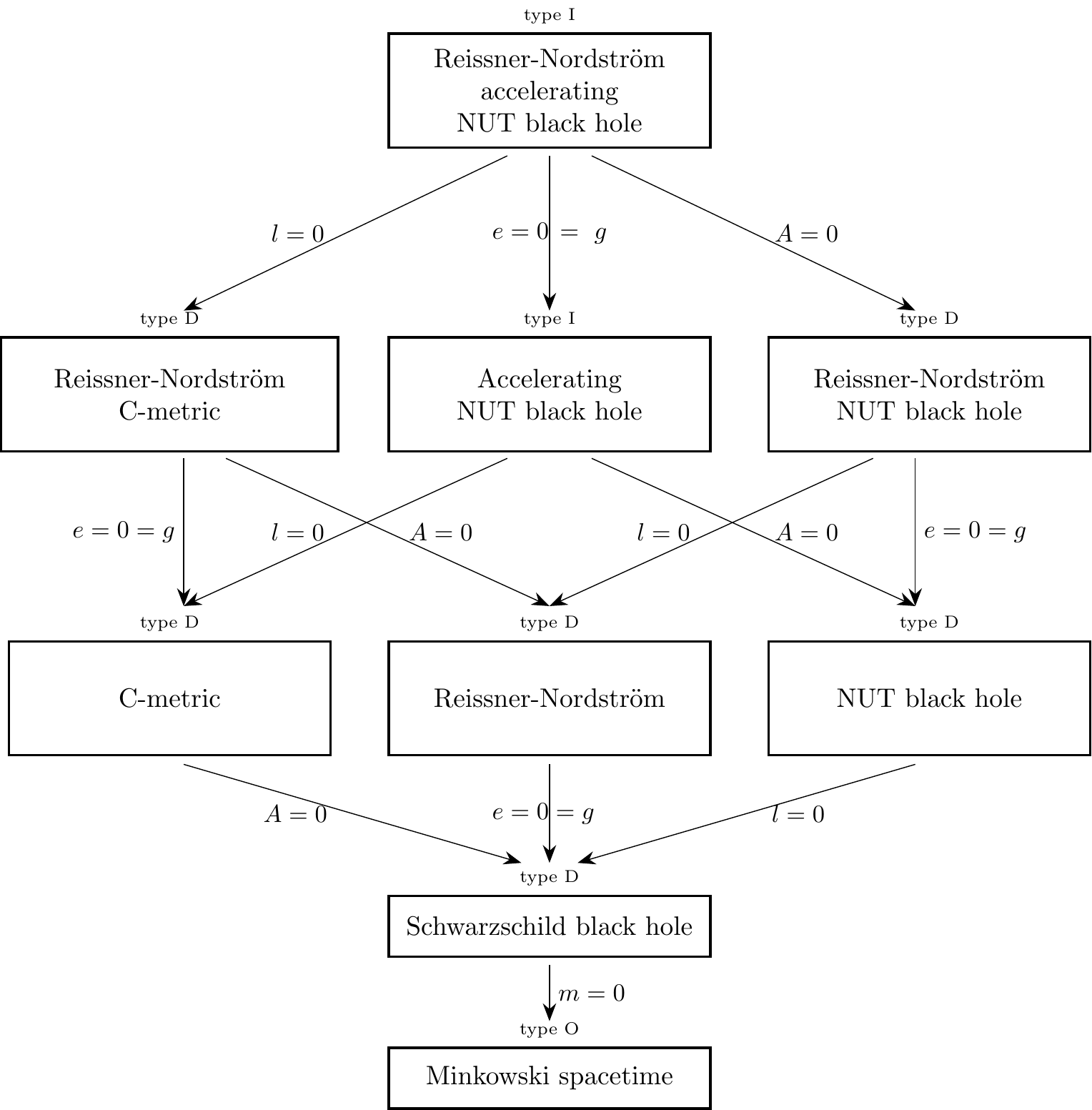}
\end{center}
\caption{Hierarchy of solutions contained in the Reissner-Nordstr\"om accelerating NUT spacetime.}
\label{fig2}
\end{figure}

In this section we tackle the construction of a NUT version of the Reissner-Nordstr\"om C-metric. As roughly mentioned before, the presence of electromagnetic charges restricts the usefulness of the standard electric Ehlers transformation as a mechanism of endorsing NUT to a given spacetime.
It was proven in Ref. \cite{Astorino:2019ljy}, by adding NUT onto the Kerr-Newman solution, that the Ehlers map requires to be complemented with a duality transformation affecting the electromagnetic potential. Specifically, the Ehlers map rotates not only the mass, producing the appearance of the NUT parameter, but also the gauge vector generating its misalignment. In order to obtain a properly aligned gauge field, a further duality transformation, that can be seen as a subcase of \eqref{gauge1}, is required to act on the electromagnetic potential. Then, the Kerr-Newman-NUT solution is obtained. An enhanced Ehlers transformation, superposing the effect of the original Ehlers map and the aforementioned duality transformation, was also provided and shown to correctly add NUT to a nonaccelerating charged spacetime while simultaneously maintaining the correct alignment of the electromagnetic potential. 
Here, we shall see that complementing the usual Ehlers map with the aforementioned duality transformation  correctly produces an accelerating NUT Reissner-Nordstr\"om solution with the correct alignment of the electromagnetic potential in the corresponding limiting cases. 
To proceed, we start from the line element \eqref{cmetricseed} with metric functions
\begin{gather}
Q(R)=\left(1-A^2 R^2\right)\left(1-\frac{2 M}{R}+\frac{e^2+g^2}{R^2}\right), \\
P(\theta)=1+2 A M \cos \theta+A^2\left(e^2+g^2\right) \cos ^2 \theta,
\end{gather}
and with the electromagnetic potential
\begin{equation}
A=-\frac{e}{R} d t+g \cos \theta d \phi.
\end{equation}
We consider both electric and magnetic charges. Again, by comparison, we can easily find the seed metric functions which take the same form as those of the uncharged case \eqref{seedfunctions}.
However, due to the presence of the electromagnetic charges, we also need to identify the electromagnetic potentials composing the seed potential $\Phi_0$. This, in turn, implies the computation of the seed twisted potential $\tilde{A}_{\varphi 0}$. Using \eqref{twistaphi}, we obtain
\begin{equation}
\tilde{A}_{\varphi 0}=-\frac{g}{R}
\end{equation}
and identify the seed Ernst potentials
\begin{equation}
\mathcal{E}_0=\frac{Q}{\Omega^2}-\frac{e^2+g^2}{R^2}, \quad \Phi_0=-\frac{e+i g}{R},
\end{equation}
which by means of the corresponding Ehlers transformations deliver
\begin{equation}
\mathcal{E}=\frac{\frac{Q}{\Omega^2}-\frac{e^2+g^2}{R^2}}{\Lambda}, \quad \Phi=\frac{-\frac{e+i g}{R}}{\Lambda}.
\end{equation}
Here we have defined
\begin{equation}
\Lambda(R, \theta)=1+i c\left(\frac{Q}{\Omega^2}-\frac{e^2+g^2}{R^2}\right).
\end{equation}
As usual, the new metric functions and the new electromagnetic potential can be easily obtained by taking the real and imaginary parts of $\mathcal{E}$ and $\Phi$. For the metric functions, we have
\begin{equation}
f=\operatorname{Re} \mathcal{E}+|\Phi|^2=\frac{\frac{Q}{\Omega^2}}{|\Lambda|^2},
\end{equation}
while the imaginary part
\begin{equation}
\chi=\operatorname{Im} \mathcal{E}=-c \frac{\left(\frac{Q}{\Omega^2}-\frac{e^2+g^2}{R^2}\right)^2}{|\Lambda|^2}
\end{equation}
provides, via \eqref{eqchi}, the corresponding function $\omega$. We obtain
\begin{equation}
\omega(R, \theta)=2 c\left(2 M \cos \theta-A\left(e^2+g^2\right) \sin ^2 \theta+\frac{A P R^2 \sin ^2 \theta}{(1+A R \cos \theta)^2}\right) .
\end{equation}
On the other hand, for the gauge potentials, we observe that
\begin{align}
& A_t(R, \theta)=\operatorname{Re}(\Phi)=-\frac{\frac{e}{R}}{|\Lambda|^2}-c \frac{\frac{g}{R}\left(\frac{Q}{\Omega^2}-\frac{e^2+g^2}{R^2}\right)}{|\Lambda|^2}, \\
& \tilde{A}_{\varphi}(R, \theta)=\operatorname{Im}(\Phi)=-\frac{\frac{g}{R}}{|\Lambda|^2}+c \frac{\frac{e}{R}\left(\frac{Q}{\Omega^2}-\frac{e^2+g^2}{R^2}\right)}{|\Lambda|^2} .
\end{align}
Notice that here we have obtained $\tilde{A}_{\varphi}(R, \theta)$ and not the real magnetic field. It is then necessary to use Eq. \eqref{twistaphi} to obtain $A_{\varphi}$, for which it is first mandatory to know the explicit form of the rotation function $\omega$. The magnetic field is found to be
\begin{equation}
A_{\varphi}(R, \theta)=c \frac{e P \sin ^2 \theta}{(1+A R \cos \theta)^2 \cos \theta}+g \cos \theta-c \frac{e}{\cos \theta}-\omega A_t.
\end{equation}
At this stage, it is useful to proceed with the same reparametrization of the mass parameter and change of coordinates performed in the uncharged case provided in Sec. \ref{seciii}. The line element is given by \eqref{lineelementmann} with metric functions 
\begin{align}
\Omega(r,\theta)&=1+A(r-r_-)\cos\theta,\\
\mathcal{P}(\theta)&=1+A(r_+-r_-)\cos\theta+A^2(e^2+g^2)\cos^2\theta,\\
\mathcal{Q}(r)&=(1-A^2(r-r_-)^2)(r^2-2mr-l^2+e^2+g^2),\\
\mathcal{T}(r,\theta)&=\frac{\mathcal{P}(\theta)(r-r_-)^2}{(r_+-r_-)\Omega(r,\theta)^2}-\frac{e^2+g^2}{r_+-r_-},\\
\mathcal{R}^2(r,\theta)&=\frac{1}{r_+^2+l^2}\left[r_+^2(r-r_-)^2+l^2\frac{\left(\mathcal{Q}(r)-(e^2+g^2)\Omega(r,\theta)^2\right)^2}{\Omega(r,\theta)^4(r-r_-)^2}\right], \\
\omega(r,\theta)&=\frac{2l(r_+-r_-)}{r_+}\left(\cos\theta+A\mathcal{T}(r,\theta)\sin^2\theta\right),
\end{align}
while the corresponding electromagnetic potentials turn to be
\begin{equation}
\begin{aligned}
A_\tau(r,\theta)&=\frac{(r-r_-)^2}{\mathcal{R}^2(r,\theta)}\left(-\frac{e}{(r-r_-)}-gl\frac{(\mathcal{Q}(r)-(e^2+g^2)\Omega(r,\theta)^2)}{r_+\Omega(r,\theta)^2(r-r_-)^3}\right),\\
A_\varphi(r,\theta)&=l\frac{e\mathcal{P}(\theta)\sin^2\theta}{r_+\cos\theta\Omega(r,\theta)^2}+g\cos\theta-l\frac{e}{r_+\cos\theta}-\omega(r,\theta)A_\tau(r,\theta).
\end{aligned}
\end{equation}
It can be observed that from the line element it is direct to obtain the NUT Reissner-Nordstr\"om metric in the limit of vanishing acceleration. However, as we have pointed out before, the Ehlers transformation also rotates the vector potential, and, as a consequence, the usual gauge field of the NUT Reissner-Nordstr\"om configuration is not retrieved. It has been proposed in Ref. \cite{Astorino:2019ljy} that an extra duality rotation acting on the vector potential 
\begin{equation}
\Phi\rightarrow\bar{\Phi}= e^{i\beta}\Phi,  \label{dualitytransf}
\end{equation}
with $\beta$ a constant, fixes the problem. This was proven for the case in which the Kerr-Newman solution is affected by an Ehlers transformation in order to obtain the Kerr-Newman-NUT spacetime. 
In addition, the usual Ehlers and the previous duality transformations were combined to produce an enhanced Ehlers transformation that  automatically adds a NUT parameter onto a charged spacetime with no extra rotation of the gauge potential. 
For the sake of simplicity, here we consider $g=0$ and apply the duality transformation (\ref{dualitytransf}) in order to align the gauge field in the corresponding subcases in which either the acceleration or the NUT parameter vanishes.
Before moving to the $(\tau,r)$ coordinates and previous to the redefinition of the mass $M$, the duality transformed electric and twisted magnetic potentials are given, respectively, by  
\begin{equation}\label{atyaphitildeconbeta}
\begin{aligned}
\bar{A}_t(R,\theta)&=\Re(\bar{\Phi})=-\frac{\frac{e}{R}}{\vert\Lambda\vert^2}\cos\beta-c\frac{\frac{e}{R}\left(\frac{Q}{\Omega^2}-\frac{e^2}{R^2}\right)}{\vert\Lambda\vert^2}\sin\beta,\\
\bar{\tilde{A}}_\varphi(R,\theta)&=\Im(\bar{\Phi})=c\frac{\frac{e}{R}\left(\frac{Q}{\Omega^2}-\frac{e^2}{R^2}\right)}{\vert\Lambda\vert^2}\cos\beta-\frac{\frac{e}{R}}{\vert\Lambda\vert^2}\sin\beta,
\end{aligned}
\end{equation}
from where the dipole magnetic potential is extracted: 
\begin{equation}
\begin{aligned}
\bar{A}_\varphi&=\frac{ce}{\Omega^2}\left[e^2A^2\sin^2\theta\cos\theta+2AM\sin^2\theta-(\Omega+1)AR-\cos\theta\right]\cos\beta+e\cos\theta\sin\beta-\omega \bar{A}_t.
\end{aligned}
\end{equation}
In the $A\rightarrow0$ limit, we get that
\begin{equation}
\begin{aligned}
\bar{A}_\varphi&=-ce\cos\theta\cos\beta+e\cos\theta\sin\beta-\omega \bar{A}_t, 
\end{aligned}
\end{equation}
an expression that can be easily rearranged into the fashion $\bar{A}_\varphi=-\omega \bar{A}_t$, by suitably selecting the constant parameter $\beta$. 
Passing to $\tau$, $r$, and $m$ and rescaling the coordinates $\tau$ and $r$ and the parameters $m$, $l$, and $e$ by the constant factor $\sqrt{\frac{r_+}{r_+-r_-}}$, the electric potential reduces to 
\begin{equation}
\bar{A}_\tau=-\frac{er}{r^2+l^2},
\end{equation}
namely, to the standard NUT Reissner-Nordstr\"om gauge field.\footnote{An accelerating version of the NUT Reissner-Nordstr\"om black has been also constructed in Ref. \cite{Astorino:2023elf}.} A hierarchy of solutions of this kind is depicted in \autoref{fig2}. 

\section{Outlook}\label{secvi}

Our study leverages Ernst's insightful description of axially symmetric and stationary spacetimes to devise a novel and more computationally efficient approach for constructing accelerating NUT black holes. Specifically, we employed an electric Ehlers transformation on the conventional C-metric to rederive Chng, Mann, and Stelea's solution \cite{Chng:2006gh} while also presenting for the first time its scalar conformally dressed generalization.
In addition, we provide the Reissner-Nordstr\"om accelerating NUT black hole of Einstein-Maxwell theory, providing the charged extension of the Chng, Mann, and Stelea solution written in Podolsk\'y and Vr\'atn\'y coordinates. A duality transformation has been applied on top of the usual Ehlers map in order to restore the alignment of the gauge field in the corresponding nonaccelerating limit. As a matter of fact, the outcome of this procedure is the obtention of the proper Reissner-Nordstr\"om NUT black hole configuration in the vanishing acceleration case. Indeed, this solution is of type I algebraic nature. 
In our constructions, we have found the coordinate system introduced by Podolsk\'y and Vr\'atn\'y \cite{Podolsky:2020xkf}  to be particularly useful, as it allows for the transparent identification of the NUT charge. Moreover, it is the appropriate change of coordinates, no matter the presence of electromagnetic charges, that allows for clean nonaccelerating limits. 
Our metric in the conformally coupled case possesses clear limits to well-known subcases, including the conformally dressed C-metric \cite{Charmousis:2009cm,Anabalon:2009qt} and NUT \cite{Bhattacharya:2013hvm,Bardoux:2013swa} black holes, which are obtained by taking the limits $l \rightarrow 0$ and $A \rightarrow 0$ on Eq. \eqref{nutcmetricconformally}, respectively. These conformally dressed spacetimes, which depart from the Pleban\'ski-Demia\'nski family, could not have been obtained from the wide family of metrics presented in Ref. \cite{Anabalon:2009qt}. 

It is known that the source of acceleration in these spacetimes is given by the presence of cosmic strings or struts, which is an unavoidable consequence of a nontrivial function $P(\theta)$ causing a deficit or excess angle on the metric. To construct accelerating spacetimes with no conical singularities, it is necessary to provide an alternative mechanism for the acceleration. Therefore, it would be interesting to explore immersing these types of solutions in external magnetic fields \cite{Astorino:2013xc} or in rotating backgrounds as  has recently been done for the standard C-metric \cite{Astorino:2022prj}. This can be easily performed by using magnetic Harrison and Ehlers transformations. 
There are several other avenues that can be explored to continue studying these geometries. In the context of our charged and hairy accelerating NUT black hole, it would be desirable to study its Euclidean version and its Eguchi-Hanson formulation \cite{Eguchi:1978xp,Eguchi:1978gw}. This would allow us to explore charged and hairy accelerating gravitational solitons, an avenue that has thus far been explored only for the nonaccelerating case \cite{Barrientos:2022yoz}. Additionally, the thermodynamic analysis of these solutions offers a highly nontrivial arena for exploration. As  is known, the C-metric and NUT solutions in GR already offer interesting subtleties regarding the computation of their conserved charges and thermodynamic behavior \cite{BallonBordo:2021pzm,Clement:2019ghi,Ciambelli:2020qny,Anabalon:2018qfv}. Therefore, accelerating NUT geometries provide an interesting class of spacetimes to test standard methods for computing charges and black hole thermodynamic laws.
On the contrary, investigating the geodesic motion of accelerating NUT black holes presents a promising avenue for exploration. Recent findings have demonstrated that geodesic observers in NUT spacetimes can avoid encountering the Misner string or experiencing any causality violations, provided that the constant parameter resulting from the coordinate transformation that determines the position of the Misner string is appropriately restricted \cite{Clement:2015cxa}. This allows for reinterpretation of NUT black holes, in some cases as traversable wormhole geometries with no energy violation whatsoever \cite{Clement:2015aka,Clement:2022pjr}. Then, it would be interesting to seek for accelerating NUT traversable wormholes, especially in the charged case scenario.

Last but not least, a major challenge will be the construction of a rotating (Kerr-like) extension of these (charged) accelerating NUT spacetimes. As pointed out in Ref. \cite{vratny}, an educated guess indicates that adding rotation to these spacetimes would guide us to a different solution than the metric contained in the Pleban\'ski-Demia\'nski family. This is expectable from the fact that, contrary to the spinning C-metric with NUT charge contained in the Pleban\'ski-Demia\'nski line element, here our accelerating NUT spacetimes are not centered with respect to $r=0$ but with respect to the position of the inner horizon $r=r_{-}$. This can be observed from the $\left(r-r_{-}\right)$ dependence of the conformal factor and accelerating horizons. In addition, the nonrotating solution already departs from the wide Pleban\'ski-Demia\'nski class, and rotation is not expected to restore the special algebraic character of this spacetime. 
An analytic construction of such a spacetime would be hard to imagine without the utilization of sophisticated solution generating techniques; we expect that some light might be shed on this problem by considering the inverse scattering method \cite{Belinski:2001ph}.

\section{Acknowledgments} 

We thank Professor Ji\v{r}i Podolsk\'y for his valuable comments and suggestions.
The work of  J. B.  is supported by the Academy of Sciences of the Czech Republic (RVO 67985840), Project No. L100192101 and by FONDECYT Grant No. 3230596. A. C.'s work is funded by FONDECYT Regular Grant No. 1210500. Both authors appreciate the technical support provided by Marcel Y\'a\~nez when dealing with heavy computational calculations.

\vspace{0.5cm}



\end{document}